\documentclass[12pt,preprint]{aastex}

\newcommand{\unit}[1]{\,{\rm #1}}

\shorttitle{3D Bubble simulations}
\shortauthors{W. Liu \emph{et al.}}

\begin{document}

\title{Long Term Evolution of Magnetized Bubbles in Galaxy Clusters}

\author{Wei Liu\altaffilmark{1}, Hui Li\altaffilmark{1}, Shengtai Li\altaffilmark{1}, Scott C. Hsu\altaffilmark{2} }

\altaffiltext{1}{Theoretical Division, Los Alamos National
  Laboratory, Los Alamos, NM, USA 87545;~wliu@lanl.gov, hli@lanl.gov, sli@lanl.gov.}\altaffiltext{2}{Physics Division, Los Alamos National Laboratory, Los Alamos, NM, USA 87545;~scotthsu@lanl.gov.}
  
\begin{abstract}
We have performed nonlinear ideal magnetohydrodynamic simulations of the long term
evolution of a magnetized low-density ``bubble" plasma formed by a radio galaxy in a 
stratified cluster medium.  It is found that about 3.5\% of the initial magnetic energy
remains in the bubble after $\sim 8 \times 10^{9}$~years, and the initial magnetic bubble
expansion is adiabatic.  The bubble can survive for at least $8 \times 10^9$~years
due to the stabilizing effect of the bubble magnetic field on Rayleigh-Taylor
and Kelvin-Holmholtz instabilities, possibly accounting for ``ghost cavities" as observed
in Perseus-A\@.  A filament structure spanning about 500~kpc is formed along the path of bubble motion.
The mean value of the magnetic field inside this structure is $\sim 0.88$~$\mu$G at $\sim8\times10^9$~years.
Finally, the initial bubble momentum and rotation have limited influence on the long term evolution of the bubble. 
\end{abstract}



\section{Introduction}

An unsolved problem in active galactic nuclei (AGN) feedback on clusters is how to account for the the morphology and stability of buoyant bubbles and their interactions with the ambient intracluster medium (ICM) \citep{mn07}.
\citet{fstacji06} showed that in the Perseus cluster, such bubbles can stay intact far from cluster centers where they were inflated by AGN jets. However, studies of kinetic-energy dominated jets in the purely hydrodynamic limit did not explain the observed long term persistence of the buoyant bubbles, which are prone to Rayleigh-Taylor (RTI) and Kelvin-Holmhotz (KHI) instabilities and fragment entirely within $100\;\unit{Myr}$, contrary to observations [but see \citet{rmfs05,ps06,ga07}]. 

Appreciable magnetic energy has been observed in both cluster and radio lobe plasmas \citep{oek00,kdl01,chh05}. The magnetic fields could play a vital role in the dynamics of the rising bubble, as shown by a series of 2D magnetohydrodynamic (MHD) studies for bubbles of several $10^8$~years
\citep{bk01,rdrr04,jd05}. \citet{sg08a} showed that uniform strong magnetic fields do not suppress RTI completely, but sheared fields at the bubble interface
quench the instability. This implies that not only the field strength but also the configuration matters for bubble stability. 
\citet{rebhp07} did a comprehensive study of the influence of different magnetic field configurations upon the stability of a rising bubble. They found that the internal bubble helical magnetic field moderately stabilizes the bubble, however the bubble has an initial plasma parameter $\beta\equiv\left<2nT/B^2\right>\gg1$,  still in thermal energy-dominated regime. 
Recently, \citet{llf06} proposed a different bubble magnetic field configuration for jet/lobes. Here we adopt the field configuration of \citet{llf06} and focus on the late stage of the evolution of magnetically dominated bubbles in 3D\@. We show that this spheromak-like magnetic field configuration strongly stabilizes the instabilities and prevents 
the bubble from breaking up, possibly forming the intact but detached ``ghost cavity" observed in systems such as Perseus-A\@. This {\it Letter} is organized as follows. In Sec.~\ref{setup}, we describe the problem setup. The simulation results and discussions are given in Sec.~\ref{results}.  

\section{Problem Set Up}\label{setup}

The background ICM is assumed to be hydrostatic and
isothermal. The density and pressure profile is given as
$\rho=p=[1+(R/R_c)^2]^{-\kappa}$, where the parameters $\kappa$
and $R_c$ are taken to be $1.0$ and $4.0$ respectively.  The fixed yet distributed gravitational field $-\nabla\psi(R)$ dominated by the dark matter is assumed \citep{nll06}. We simulate
only the phase after the AGN has inflated a
low-density bubble which is in pressure equilibrium
with the ICM\@ initially.  The static magnetized bubble initially lies at
$(x_b,y_b,z_b)=(0,0,5)$ with density $\rho_b=0.1$, spherical radius
$r_d=2$ for the density profile and $r_b=1$ for the magnetic
configuration. 
The specific heat $\gamma$ inside and outside the bubble is taken to be $5/3$. Physical
quantities are normalized by the characteristic system length scale
$R_0=25\;\unit{kpc}$, density
$\rho_0=1.67\times10^{-26}\;\unit{g\;cm^{-3}}$, and velocity
$V_{0}=6.2\times10^{7}\;\unit{cm\;s^{-1}}$. The initial sound speed,
$C_s|_{t=0}=\gamma^{1/2}\sim1.29$, is constant throughout the
computational domain. Other quantities are normalized as: time $t=1$ gives
$R_0/V_0=38.6\;\unit{Myr}$, magnetic field $B=1$ gives
$(2\rho_0V_{0}^2)^{1/2}=1.13\times10^{-5}\;\unit{G}$ and energy $E=1$
gives $\rho_0
V_{0}^2R_0^3=2.71\times10^{58}\;\unit{ergs}$. The magnetic field setup
of the bubble follows \citet{llf06} with the ratio of toroidal to
poloidal bubble field $\alpha$ taken to be $\sqrt{10}$, which
corresponds to a minimum initial Lorentz force.  The latter is
reasonable since we want to mimic the situation that the magnetic
bubble has substantially relaxed and detached from the jet tip. The
initial total magnetic energy is around
$2.0\times10^{59}\;\unit{ergs}$.  The computational domain is taken to
be $|x|\le16$, $|y|\le16$, and $0\le z\le32$ corresponding to a
$(800\;\unit{kpc})^3$ box in the actual length scales. The numerical
resolution used here is $400^3$, where the grid points are assigned
uniformly in every direction. A cell $\delta x$ corresponds to
$2.0\;\unit{kpc}$. We use ``outflow" boundary conditions at every
boundary except the boundary at $z=0$, where we use ``reflecting"
boundary conditions, which guarantee that the energy flux through this
boundary is zero.

\section{Results \& Discussion}\label{results}

\begin{figure}[!htp]
\begin{center}
\scalebox{0.2}{\includegraphics{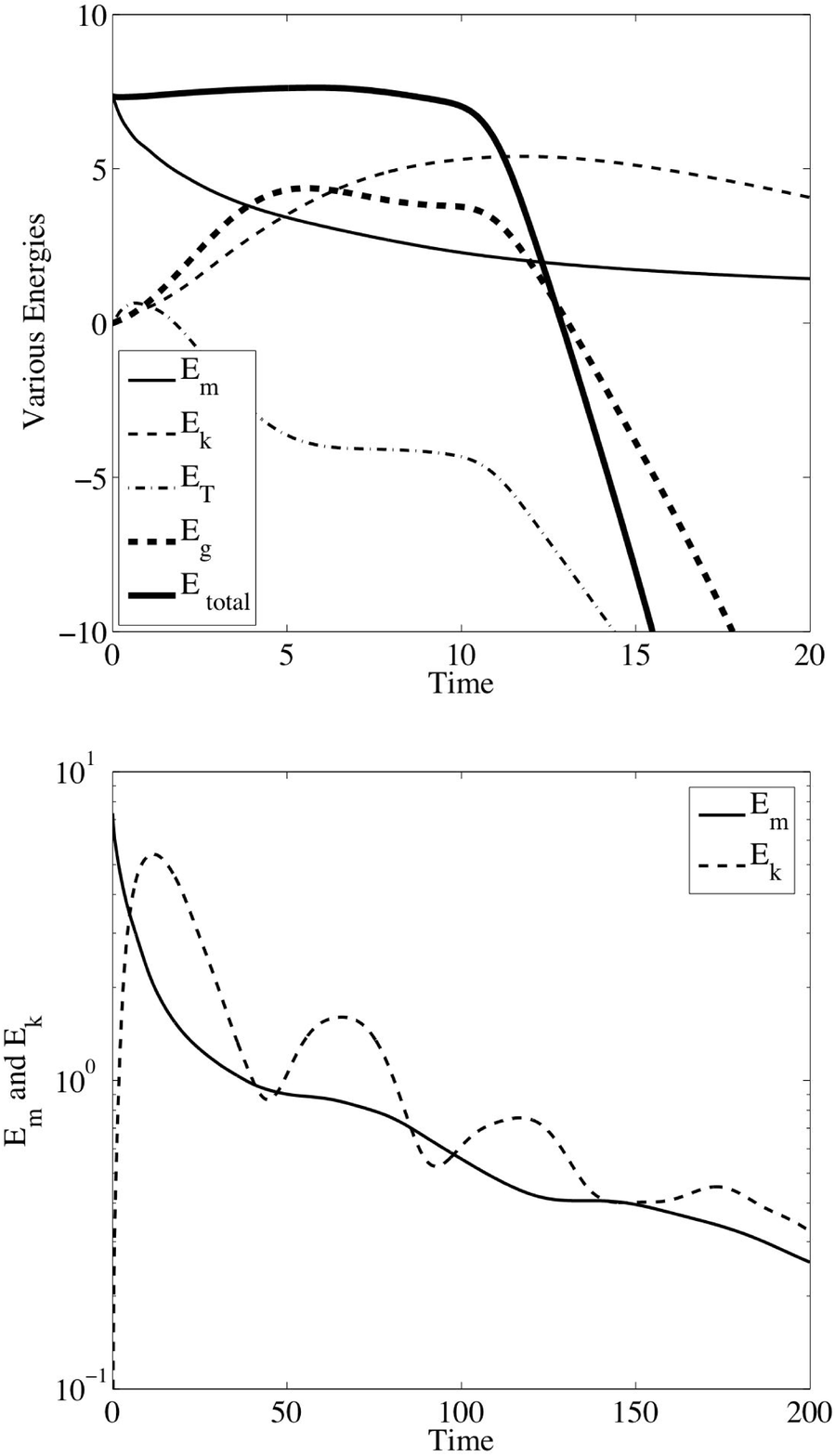}}
\caption{\label{energy}~Time evolution of various  energies. Top: short term various energies evolution $t\le20$, thin solid: magnetic Energy $E_m$, thin dash: kinetic energy $E_k$, thin dash dot: internal energy $E_T$, thick dash: gravitational energy $E_g$ and thick solid: total energy $E_{\rm tol}$. Bottom: long term time evolution of magnetic energy (in logarithmic scale) $E_m$ (solid) and kinetic energy $E_k$ (dash). $t=1$ represents $38.6\;\unit{Myr}$.}
\end{center}
\end{figure}

Energy and density evolution reveal the different stages of the bubble. Figure~\ref{energy}(\emph{top})
presents the time evolution of various  energies at the early stage,
in which the gravitational energy $E_g=\int \rho\psi dV$ and the
internal energy $E_T=\int p/(\gamma-1)dV$, where $dV$ is the
infinitesimal volume and the integral is over the entire computation
domain, have been reduced by their
initial values $E_{g,0}$ and $E_{T,0}$, respectively. The total energy
is defined as $E_{\rm total}=E_m+E_k+E_T+E_g$, where the  kinetic
energy is defined as $E_k=\int 1/2\rho v^2 dV$ and the  magnetic
energy is defined as $E_m=\int B^2/2 dV$. 
At the early stage ($t\lesssim5$), the kinetic energy increases since
the bubble accelerates upward due to buoyancy, and the magnetic energy 
decreases due to work done in expansion from the weak initial Lorentz force and conversion to shock and wave energy \citep{nll06}.
The gravitational energy increases because there is a net outward mass flow in the axial direction.
The passage of the shock wave heats and compresses the ICM and alters
its pressure gradient.  The total energy $E_{\rm total}$ is almost
constant before $t\sim10$, at which time the shock and wavefront
reach the boundary. Conservation of
the total energy for $t\lesssim10$ is not strictly satisfied due to
numerical diffusion.  The initial bubble expansion ($t\le20$) is approximately adiabatic, which gives $E=Vp\propto V^{1-\gamma}$, where $V$ is the bubble volume. Therefore
$E_2/E_1=(V_1/V_2)^{2/3}$, where the subscript $1$ and $2$ indicate
the initial and final state, respectively. At $t=20$, the radius of
the bubble has increased from $2$ to $\sim5$, which gives
$E_2/E_1\sim16\%$, roughly matching the amount of magnetic energy in
the bubble [$20\%$ from Fig.~\ref{energy}(\emph{top})].  For later times
[Fig.~\ref{energy}(\emph{bottom})], fitting with
$E_m(t)/E_{m,0}=\exp(-t/\tau_{\rm dis})$, we have: (1)~a fast
dissipation stage ($t\lesssim10$) when magnetic energy
dissipation time $\tau_{\rm dis}\sim11$; (2)~a slow dissipation stage
($t\gtrsim10$) when $\tau_{\rm dis}\sim114$. Both times are much smaller
than the numerical dissipation time $\tau_{\rm res}$ at the
corresponding stages (see discussions at the end of this section). The kinetic energy is oscillating while it is
decaying slowly, which can be understood as follows.
Gravity pulls the bubble down to the denser ICM, compressing
the bubble and causing an increase in magnetic energy (since
the magnetic flux is nearly conserved).  This results
in the magnetic energy oscillating in phase with
the kinetic energy with period $\sim90$
($3.5\times10^9\;\unit{yr}$).  
At $t=200$ after
several periods of decaying oscillations, about 3.5\% of the initial magnetic energy remains.

\begin{figure}[!htp]
\begin{center}
\scalebox{0.8}{\includegraphics{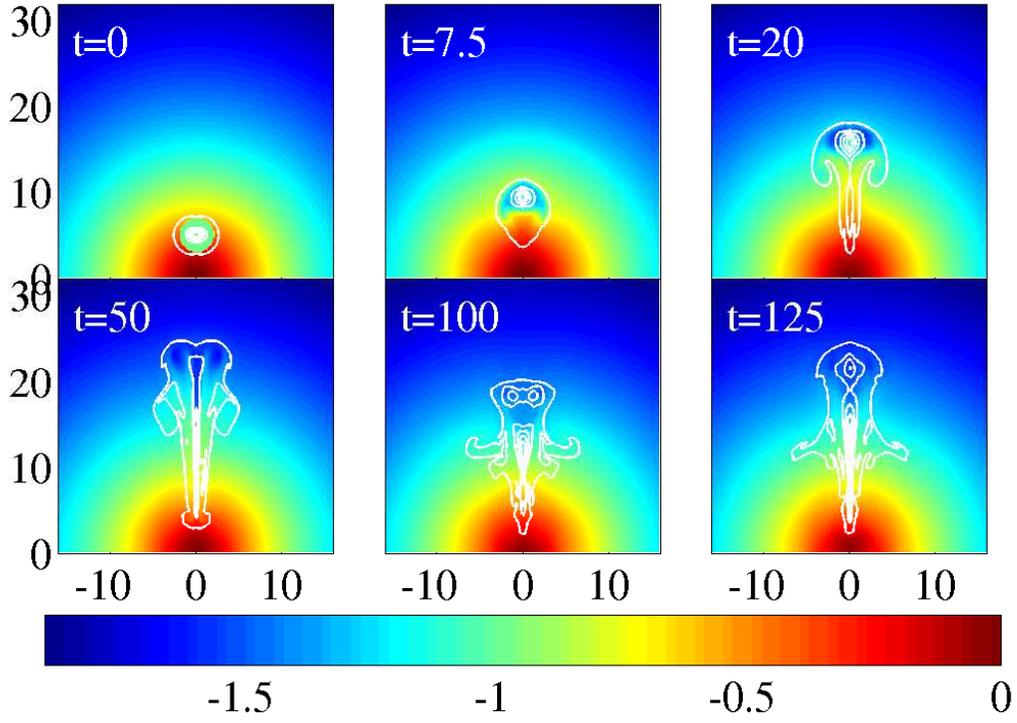}}
\end{center}
\caption{~(color) Density (in logarithmic scale) in the $x$-$z$ plane as a function of time ($\alpha=\sqrt{10}$). The white solid contour lines indicate the magnetic field strength $|B|$. $t=0$, $|B|\in[0,1.976]$; $t=7.5$, $|B|\in[0,0.830]$; $t=20$, $|B|\in[0,0.400]$; $t=50$, $|B|\in[0,0.328]$; $t=100$, $|B|\in[0,0.159]$; $t=125$, $|B|\in[0,0.156]$. $|B|=1$ represents $11.3\;\unit{\mu G}$. The number of the contour levels are all $5$. \label{overview} }
\end{figure}

\begin{figure}[!htp]
\begin{center}
\scalebox{0.4}{\includegraphics{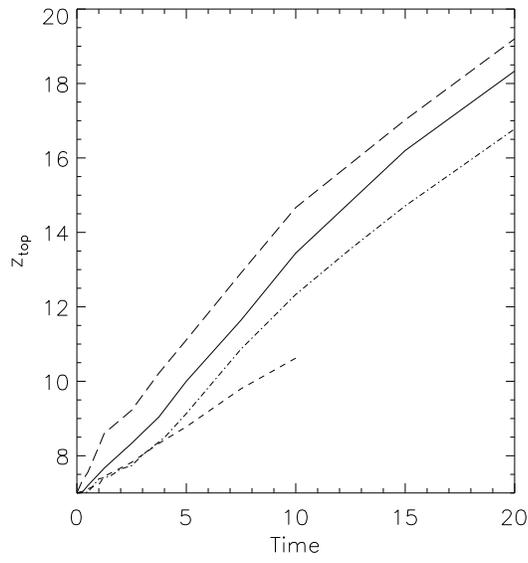}}
\end{center}
\caption{~Position of the bubble top \emph{v.s.} time for four different runs. The solid, long dash and dash dot lines represent magnetized bubble runs without initial momentum rotation, with $v_{\rm inj}=1.174$ and with $\omega=0.9675$ respectively. The short dash line denotes the purely hydrodynamic run. \label{pos} }
\end{figure}

The magnetic field suppresses instabilities and therefore the bubble
remains intact longer. Figure~\ref{overview} presents the typical
density distribution (logarithmic scale) in 2D $x$--$z$ slices at
$y=0$ at different times ($t=0,7.5,20,50,100,125$). The white solid
contour lines indicate contours of constant magnetic field strength
$|B|$. Consistent with
Fig.~\ref{energy}(\emph{bottom}), after $t\gtrsim50$, the bubble
undergoes a slowly decaying oscillation between $z\sim19$ and
$z\sim24$. We have also performed simulations of an unmagnetized
bubble and found that it disintegrates after $t=20$, whereas the
magnetic bubble still clearly differentiates itself from the ambient
medium at $t=200$ (Fig.~\ref{overview}). The formation of an ``umbrella" or a thin protective
magnetic layer on the bubble working surface suppresses
instabilities \citep{rebhp07}.  The stronger magnetic field is also
found to move the bubble faster and push the bubble farther away from
its initial position.  The position of the bubble top as a function of
time is shown in Fig.~\ref{pos}. This position is calculated by
plotting the axial profile of the density along the line of
$(x,y)=(0,0)$ and finding the location of the first density jump from
the top of the domain. Comparing the unmagnetized run (dash line) with
the magnetized run (both without initial momentum/rotation), we can
see that the rising speed increases from $~0.36$ to $0.6$.  Note that
the rough estimate of the terminal speed $v_t$ based on the initial gravitational
acceleration $g$ and size $r_d$, both position-dependent, is $v_t\approx4/3\sqrt{2gr_d}\sim2$,
assuming force balance between buoyancy and viscous drag force at the
final stage \citep{mn07}.

\begin{figure}[!htp]
\begin{center}
\scalebox{0.2}{\includegraphics{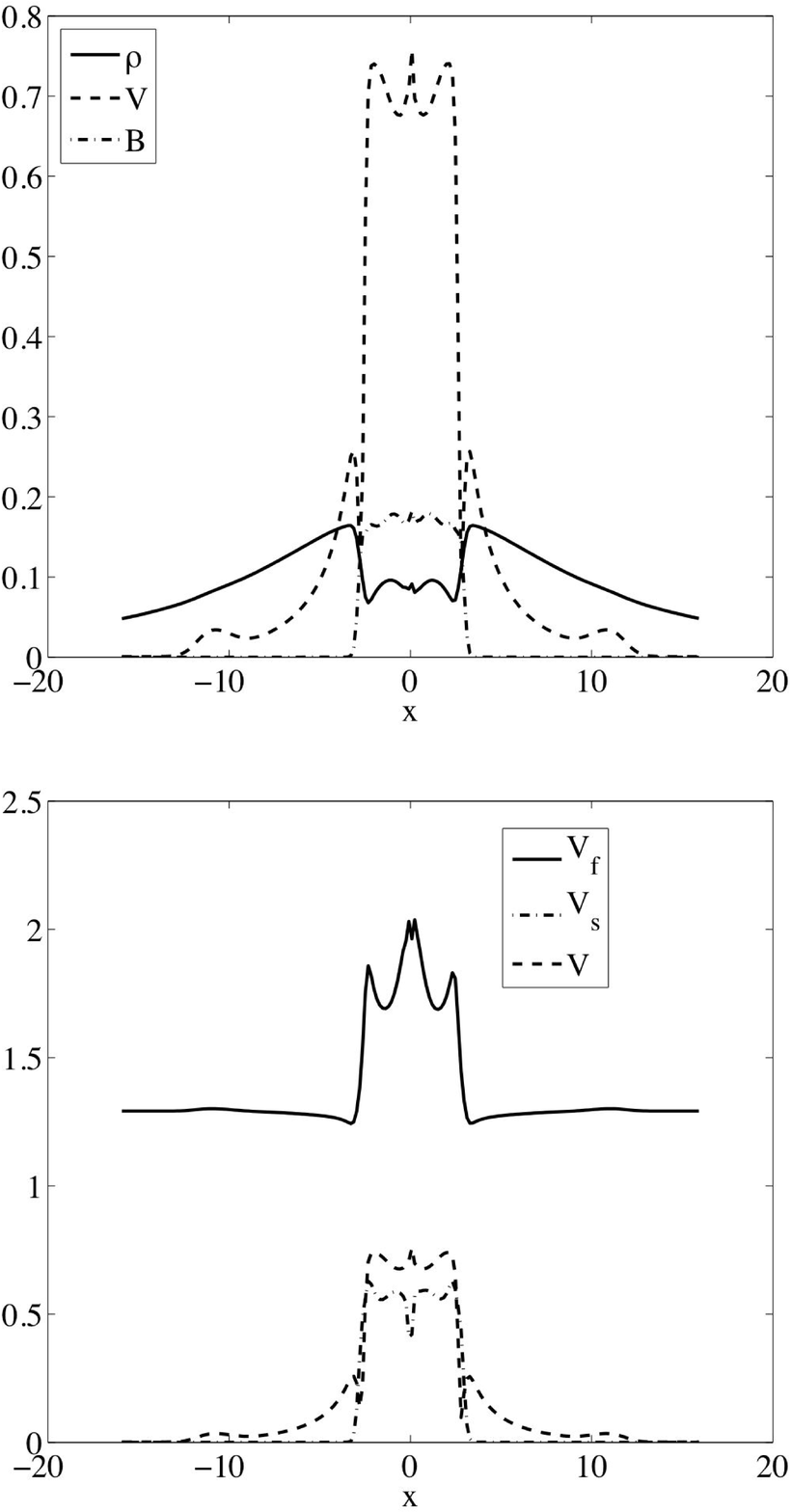}}
\caption{\label{khi}~Transverse profiles in the $x$-direction of several quantities at $t=7.5$ on $z=8.0$. The bulk speed $V$, the density $\rho$, and the magnetic field strength $B$ are shown for inspecting the KHI surface modes (\emph{top}). The bulk speed $V$, the superfast magnetosonic speed $V_f$, and the slow magnetosonic speed $V_s$ are shown for inspecting the KHI body modes (\emph{bottom}).}
\end{center}
\end{figure}

We investigate the effect of the magnetic field on bubble stability
in more detail.  Following \citet{nll07}, we first study KHI of the
bubble. At both sides of the bubble, the magnetic field is almost uniform, not twisted like at the top of the bubble. From linear analysis, the instability criterion for nonaxisymmetric KHI surface
modes is \citep{hr02} : $\Delta V > V_{As}=[(\rho_b+\rho_e)/(4\pi
\rho_b\rho_e)(B_b^2+B_e^2)]^{1/2}$, where $\Delta V\equiv|V_b-V_e|$ is
the velocity shear and $V_{As}$ is the surface Alfv\'en
speed. The subscripts $b$ and $e$ indicate the bubble and external medium,
respectively. The nonaxisymmetric body modes would be important if
\citep{hr99}: (1) $V_b>V_f$, in which $V_f$ is the fast
magnetosonic speed; or (2) $C_sV_A/(C_s^2+V_A^2)^{1/2}<V_b<V_s$, in
which $V_A$ is the Alfv\'en speed and $V_s$ is the slow magnetosonic
speed. The definitions of $V_{f,s}$ are:
$V_{f,s}=\big\{1/2\{C_s^2+V_A^2\pm[(C_s^2+V_A^2)^2-4C_s^2V_A^2\cos^2\theta]^{1/2}\}\big\}^{1/2}$.
Since we focus on the $x$-direction only, $V_A^2\cos^2\theta$ are
taken to be $B_x^2/\rho$.
Figure~\ref{khi} (\emph{top}) displays the transverse distribution of
the bulk flow speed $V=(v_x^2+v_y^2+v_z^2)^{1/2}$, the density $\rho$,
and the magnetic field strength $B$ at $t=7.5$ at $z=8.0$.
A distinct velocity shear is identified at $x\sim3.5$. Across this
shear, $\Delta V\sim0.75$ and $V_{As}\sim0.85$. The inequality $\Delta
V<V_{As}$ holds. This means that the KHI surface modes are completely
suppressed. Figure~\ref{khi} (\emph{bottom}) displays the transverse
distribution of $V$, $V_f$, and $V_s$ at $t=7.5$ at $z=8.0$ . We see
that the bulk flow $V$ lies between $V_f$ and $V_s$. The inequality
$V_s<V<V_f$ is satisfied in the body of the buoyant bubble. This
rules out the KHI body modes as well.

\begin{figure}[!htp]
\begin{center}
\scalebox{0.4}{\includegraphics{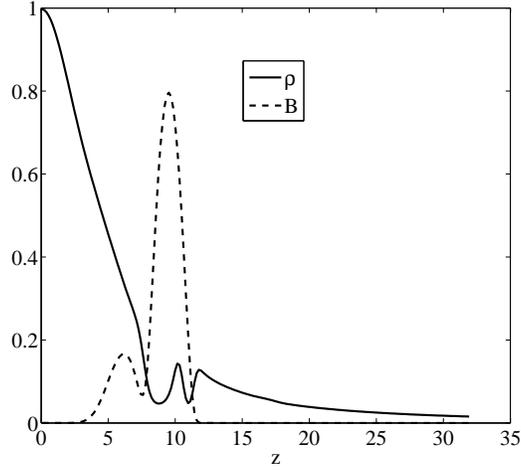}}
\caption{\label{rti}~Axial profiles in the $z$-direction of several quantities at $t=7.5$ with $(x,y)=(0,0)$. The density $\rho$ and the magnetic field strength $B$ are shown for inspecting the RTI modes.}
\end{center}
\end{figure}

RTI could also be suppressed by the magnetic field. For the idealized
case of two conducting fluids separated by a contact discontinuity
with a uniform magnetic field parallel to the interface undergoing
constant acceleration $g$, \citet{chan61} demonstrated that RTI on a
scale $L$ parallel to the field requires
$B<B_c\equiv[Lg(\rho_h-\rho_l)]^{1/2}$, in which $\rho_h$ and $\rho_l$
are the densities in the heavy and light fluids respectively. Modes
perpendicular to the field are unaffected.  At the top of the bubble
($z\sim11.6$) (Fig.~\ref{rti}), $\rho_h$ and $\rho_l$ are found to be
$0.15$ and $0.06$, respectively, and the gravitational acceleration
$g$ is calculated to be $0.95$. If $L$ is chosen to be the computation
domain size, the maximum possible mode wavelength in the simulations,
then the critical magnetic field strength is $B_c\sim1.6$, which is
larger than the magnetic field  ($\sim0.8$) at that location. This
means that only part of the parallel modes are suppressed regardless
of the perpendicular modes. This is not consistent with the simulation
results. However, as pointed out in \citet{sg08a}, the twisting nature
of the bubble field at the top of the bubble introduces a current
sheet at the surface, and the changes in the direction of the field at
the interface must be on very small scales to inhibit the interchange
modes. A more detailed study of this effect is beyond the scope of
this paper and will be the subject of future study.

Our simulation also provides one possible explanation of the
morphology and origin of the large scale magnetic field and
the generation
of ``ghost cavities" observed in many clusters. Like \citet{rdrr04}, a
magnetized high-density tail remains as the bubble rises
(Fig.~\ref{overview}). The tails have an elongated morphology
(Fig.~\ref{overview}), resembling H$\alpha$ filaments, which is found
to indicate the history of the rising bubble.  Interestingly, the
magnetic field also helps stabilize this ``filament" as it is still
visible at $t=200$.  \citet{nb04} has argued that thermal conduction
has to be strongly suppressed in the ICM; otherwise, such cold filaments
would be rapidly evaporated. As in \citet{rebhp07}, our results
provide a possibility that thermal conduction may be locally weaker in
the bubble wake, thus preventing or slowing down filament
evaporation. The magnetic field is distributed between $z\sim2.5$ and
$z\sim25$ at $t=200$ with peak value around $0.156$ ($\sim1.8\;\unit{\mu
G}$) or mean value around $0.078$ ($\sim0.88\;\unit{\mu G}$), close to
the estimates of wider cluster fields \citep{ct02,tfa02}. This
large-scale ($\gtrsim500\;\unit{kpc}$) magnetic field structure also
mimics the morphology of the second class (``Phoenix") of ``radio
relics" elongated from the cluster center to the periphery observed in
A 115 \citep{gfg01}.   The simulation results mean that the magnetic
bubble rising from $\sim125\;\unit{kpc}$ with an initial magnetic
energy of $2\times10^{59}\;\unit{ergs}$, will spread magnetic fields
between $62.5\;\unit{kpc}\lesssim z\lesssim600\;\unit{kpc}$ and keep a
magnetized bubble between $475\;\unit{kpc}$ and $600\;\unit{kpc}$ for
at least $\sim8\times10^{9}\;\unit{yr}$. Therefore one could reproduce
the intact but detached ``ghost cavity" observed in some systems such
as Perseus-A.

Real radio lobes possess complex, jet-driven flows, different from the
static bubble that we use for our initial conditions.  We study the
effects of additional variations in the initial bubble by having:
(1)~an initial uniform injection speed $v_{\rm inj}$ or (2)~an initial
uniform rotation $\omega$. \citet{ntll08} reported that the radio lobe
gains a momentum of $v_{\rm inj}\sim0.91C_{s0}=1.174$, and the
rotation speed $\omega$ at the edge of the bubble reaches
$\sim1.5C_{s0}/r_d=0.9675$ after jet-driven inflation and interaction
with the ambient ICM, although neither are not uniform inside the
bubble. Here we take these two values as our initial values for
$v_{\rm inj}$ and $\omega$ and idealize both to be uniform throughout
the bubble.  Fig.~\ref{pos} presents the time evolution of the
position of the bubble top with initial uniform injection velocity
$v_{\rm inj}=1.174$ (long dash) and initial uniform rotation speed
$\omega=0.9675$ (dash dot). It is seen that these two variables only
have some influence on the evolution of the bubble during the early
stage. The bubble with an initial rotation has a reduced buoyancy resulting from rotation-driven instabilities that in turn lead to a reduced pressure differential between the bubble and the ICM.
The evolution,
and structure of the bubble in the later times are,
however, essentially the same. 
 
The effects of numerical diffusion are estimated as follows. If we fit
the net toroidal magnetic flux $\psi_t=\int B_y dS$ (only positive $B_y$ is selected) as
$\psi_t(t)/\psi_t(t=0)\equiv\exp(-t/\tau_{\rm res})$, the ``resistive"
dissipation time due to numerical diffusion is $\tau_{\rm
res}\sim 38$ for $t\le10$, $\tau_{\rm res}=138$ for
$10\le t\le50$, and $\tau_{\rm res}=507$ for $50\le t \le
100$. Therefore the numerical diffusion is not important on the
time scales of interest $(t\lesssim200)$. Increasing numerical resolution can further reduce the numerical diffusion.

The oldest bubbles known so far are a few hundred $\unit{Myr}$ old \citep{mn07}, which are much younger than the long-sustained ($\sim$ a Hubble time) magnetized bubbles reported in this \emph{Letter}. The simulations show that at later times $t\gtrsim50$, the density and X-ray luminosity contrast between the bubble and ICM
plasma at the periphery of the cluster is small, which makes the direct X-ray observation of the bubble faraway from the parent galaxy very difficult. But the radio bubble at the borders of the clusters may still be detectable through: (1) Faraday rotation measurement of the outskirts of the cluster if an outside synchrotron source exists; (2) or radio observation from re-energization of fossil radio bubble plasma by shocks or mergers. It is highly possible for the bubbles to pile up in the outer atmospheres of the clusters in deep Chandra observation, if those observations are technically viable. 

It is conceivable that part of the initial magnetic energy of the bubble
accelerates cosmic rays.
Throughout the lifetime of a cluster, there could be multiple AGNs
injecting jets/lobes into the ICM\@.  Both the cosmic rays
and the remaining magnetic fields (distributed over large scales)
could provide the energy sources for phenomena such as radio relics and
radio haloes that have been observed for a number of clusters \citep{fgsbr08}. 



%

\end{document}